\documentclass[aps,prl,preprintnumbers,groupedaddress]{revtex4}
\newcommand{\bea}{\begin{eqnarray}}
\newcommand{\eea}{\end{eqnarray}}

\usepackage{graphicx}
\begin{document}
\preprint{Chin.Phys.Lett.{\bf 23}(2006)2688}
\preprint{nucl-th/0609068}
\title{Pad\'e expansion and the renormalization of nucleon-nucleon scattering}
\author{YANG Ji-Feng\footnote{jfyang@phys.ecnu.edu.cn, corresponding author
}, HUANG Jian-Hua, LIU Dan}
\address{Department of Physics, East China Normal University, Shanghai
200062, China}
\begin{abstract}
The importance of imposing physical boundary conditions on the
$T$-matrix to remove the nonperturbative renormalization
prescription dependence is stressed and demonstrated in two
diagonal channels, $^1P_1$ and $^1D_2$, with the help of Pad\'e
expansion.
\end{abstract}
 \maketitle
Weinberg's seminal works\cite{weinEFT} marked the advent of the
effective field theory (EFT) methods in the studies of nucleon
systems\cite{Epelrev}. However, the nonperturbative character
complicates the renormalization of such EFT, which has led to many
publications discussing this topic, see, e.g.,
Refs.\cite{vK,Epel,Cohen,Rho,KSW,Gege,Soto,BBSvK,Entem,VA,PRC71}.
The consensus now arrived is that, for such EFT's, the power
counting rules are strongly intertwined with the regularization
and renormalization prescriptions. Therefore, a number of power
counting schemes have been proposed\cite{KSW,Soto,BBSvK,Entem,VA}.

On the other hand, the strong interplay between the power counting
and renormalization makes the $T$-matrix develop a nontrivial
prescription dependence\cite{PRC71}, unlike the perturbative
cases\cite{scheme}. Such phenomenon is already known in other
nonperturbative problems\cite{YangPRD02}. Then the crucial task is
to pin down the appropriate counter terms\cite{Rho,BBSvK}, the
cutoff scales\cite{Epel}, or whatever parameters\cite{VA} in the
renormalized $T$-matrix in a fashion that the physical boundary
conditions, like phase shifts, are fulfilled.

In this short report, we wish to further demonstrate and stress
our point\cite{PRC71} in the diagonal channels, $^1P_1$ and
$^1D_2$. We will employ the Pad\'e expansion of a compact
parametrization of the $T$-matrix proposed before\cite{invT} to
make our main points about prescription dependence relatively
transparent and simple.

The basic framework for describing the nucleon-nucleon (NN)
scattering processes at low energies is the $T$-matrix that
satisfies the Lippmann-Schwinger (LS) equation in partial wave
formalism,
\begin{eqnarray}
\label{LSE} &&T_{ll^\prime}(p^{\prime},p;
E)=V_{ll^\prime}(p^{\prime},p)+\sum_{l^{\prime\prime}}\int
\displaystyle\frac{kdk^2}{(2\pi)^2} V_{l
l^{\prime\prime}}(p^{\prime},k ) G_0(k;E^+)
T_{l^{\prime\prime}l^\prime}(k,p; E), \nonumber \\
&&G_0(k;E^+)\equiv \frac{1}{E^{+}-k^2/(2\mu)},\ \ E^{+}\equiv E +
i \epsilon,
\end{eqnarray}
with $E$ and $\mu$ being respectively the c.m. energy and the
reduced mass, ${\bf p}$ (${\bf p}^\prime$) being the momentum
vector for the incoming (outgoing) nucleon, and $p^{\prime}=|{\bf
p}^\prime|$, $p=|{\bf p}|$. The potential $V(p^{\prime},p)$ can be
systematically constructed from the $\chi$PT according to
Weinberg's proposal\cite{WeinEFT}. We remind that the constructed
potential is understood to be finite first, as the 'tree' vertices
in the usual field theory terminology.

To see the nonperturbative feature, we could transform the above
LS equation into the following compact form as a nonperturbative
parametrization of $T$-matrix (we drop all the subscripts for
simpleness)\cite{invT}
\begin{eqnarray}
\label{npt} &&T^{-1}=V^{-1}-{\mathcal{G}},\\
&&{\mathcal{G}}\equiv V^{-1}\left \{\int \frac{kdk^2}{(2\pi)^2}V
G_0T\right \} T^{-1}.
\end{eqnarray}Here $V$, $T$ and therefore
${\mathcal{G}}$ are momentum-dependent matrices in angular
momentum space (to show $T T^{-1}=T^{-1} T=I$, we note that,\bea
 T T^{-1}=&&(V+V\otimes G\otimes T)T^{-1}=V
T^{-1}+V{\mathcal{G}}=V(T^{-1}+{\mathcal{G}})=VV^{-1}=I,\nonumber\\
T^{-1}T=&&V^{-1}T-{\mathcal{G}}T=V^{-1}T-V^{-1}V\otimes G\otimes
T=V^{-1}(T-V\otimes G\otimes T)=V^{-1}V=I,\nonumber\eea with
$\otimes$ denoting the convolution operation). Unfolding the
${\mathcal{G}}$ factor, one could readily see that it is an
intrinsic nonperturbative factor. The renormalization of the
$T$-matrix is to render this factor free of divergences.

In Ref.\cite{invT}, it was argued that the perturbative
subtraction programme simply fails to renormalize the
nonperturbative $T$-matrix. In Ref.\cite{PRC71}, this point was
further illustrated by rigorous solutions based on contact
potentials, where the key conceptual issues were clarified. For
example, for contact potentials, the $T$-matrix for $^1S_0$
channel would take the following compact and hence nonperturbative
form, \bea \label{1s0}\frac{1}{T^{^1S_0}(p)}=\frac{\sum_i
N^{^1S_0}_i p^{2i}}{\sum_j D^{^1S_0}_j p^{2j}}+i\frac{\mu
p}{2\pi}, \ p=\sqrt{2\mu E},\eea where the parameters
$[N^{^1S_0}_i,D^{^1S_0}_j]$ are functionally dependent upon the
EFT couplings (appearing in the potential) and the renormalization
prescription parameters (counter terms, cutoffs, subtraction
points, etc.).

Thus, to successfully renormalize the nonperturbative
$\mathcal{G}$ or $T$-matrix, the counter terms must be introduced
in such a fashion that each of the parameters
$[N^{^1S_0}_i,D^{^1S_0}_j]$ is finite. In general cases, it means
that the counter terms must be introduced {\em before} the
corresponding infinite perturbative series are summed up or {\em
before} the corresponding Schr\"odinger equation is
solved\cite{BBSvK}. Such counter terms has been termed to be
'endogenous'\cite{invT,PRC71} to stress the difference from the
perturbative cases.

Before removing the divergences, the ${\mathcal{G}}$ factor in a
diagonal channel would essentially take the following form \bea
\text{Re}({\mathcal{G}}_l(p))= \frac{\sum N^{(Bare)}_i(\Lambda,
[C_k]) p^{2i}+\text{finite pieces(nonlocal)}}{\sum D^{(Bare)}_j
(\Lambda, [C_k]) p^{2j}+\text{finite pieces(nonlocal)}}, \eea with
$\Lambda$ being a cutoff scale and $[C_k]$ being the EFT
couplings. This is because the divergences brought about with the
potential constructed from $\chi$PT are mainly power divergences
(contained in $[N^{(Bare)}_i, D^{(Bare)}_j ]$). In the relatively
low energy regions for the $NN$ scattering, say $E\in (0,
200\text{MeV})$, such power like terms dominates the
${\mathcal{G}}$ factor. Therefore, after subtracting the
divergences through endogenous counter terms, the nontrivial
prescription dependence in  the ${\mathcal{G}}$ factor (and hence
in the $T$-matrix) is universally parametrized in terms the
renormalized constants $[N^{(r)}_i, D^{(r)}_j ]$.

This discussion points towards a simple technical treatment of the
general parametrization of the nonperturbative prescription of the
$T$-matrix: Parametrize ${\mathcal{G}}$ in the low energy regions
via Pad\'e (or Taylor, in  even lower regions) expansion: \bea
\label{PADE} &&\text{Re}({\mathcal{G}}_l (p))|_{\text{Pad\'e}}=
\frac{n_{l;0}+n_{l;1}p^2+\cdots}{d_{l;0}+d_{l;1}p^2
+\cdots},\\
&&\text{Re}({\mathcal{G}}_l (p))|_{\text{Taylor}} = g_{l;0}
+g_{l;1} p^2 +\cdots. \eea Here the coefficients
$[n_{l;i},d_{l;j}]$ or $[g_{l;n }]$ will inevitably be
renormalization prescription dependent. Interestingly, such
treatment leads to a general parametrization of the
renormalization prescription. Now it is straightforward to see the
utility of this simple treatment in the following aspects: 1) Much
labor is spared in solving the LS equation with various
approximation methods and the subsequent renormalization of the
$T$-matrix, which often proves difficult; 2) One avoids being
stuck in a specific renormalization prescription when there is a
need to examine the prescription sensitivity of the conclusions
obtained (This point has been overlooked and caused some disputes
on some important issues\cite{chirallimit}.); 3) One could test
whether EFT systematically works for $NN$ scattering in a simple
manner. In fact more virtues could be enumerated.

Now our main point becomes obvious: Since the $T$-matrix must
yield the physical predictions, like phase shifts, while different
constants of $[n_{l;i},d_{l;j}]$ or $[g_{l;n }]$  or different
prescriptions will give different predictions for the phase
shifts, only one set of values for these constants (up to
equivalence) could correspond to physical situation. In fact, even
if one works with a rational power counting system, the
predictions could not be relevant to physical situation {\em if}
the renormalization prescription in use is not fixed by physical
boundary conditions. One might doubt that the coarse treatment
described about would be useful in practice even if our viewpoint
is correct. In the following, to show the efficiency of this
seemingly coarse treatment, we demonstrate the phase shift
predicted by using such Pad\'e expansion for two diagonal
channels, $^1P_1$ and $^1D_2$. For the $^1S_0$ channel case,
please see Ref.\cite{1s0}.


The strategy is as follows: 1) First, we choose the truncation of
the potential and Pad\'e expansion (or Taylor expansion); 2)
Second, we fit to the phase shift data in the low energy ends, say
$E(=\text{T}_{\text{lab}}\ \text{in the figures})\in (0,
10\text{MeV})$, to determine the coefficients (which represent
prescription parametrization) in the expansion; 3) Third, the
phase shift curves in remaining regions, say, $E\in
(10\text{MeV},200\text{MeV})$, are predictions. Obviously, the
second step is crucial, and corresponds to the step of imposing
physical boundary conditions to fix counter terms, cutoff scales
or their equivalents in conventional
approaches\cite{Epel,Rho,KSW,BBSvK,Entem,VA}.

\begin{figure}[t]
\begin{center}
\begin{tabular}{cc}
\hspace{-0.4cm} \resizebox{90mm}{!}{\includegraphics{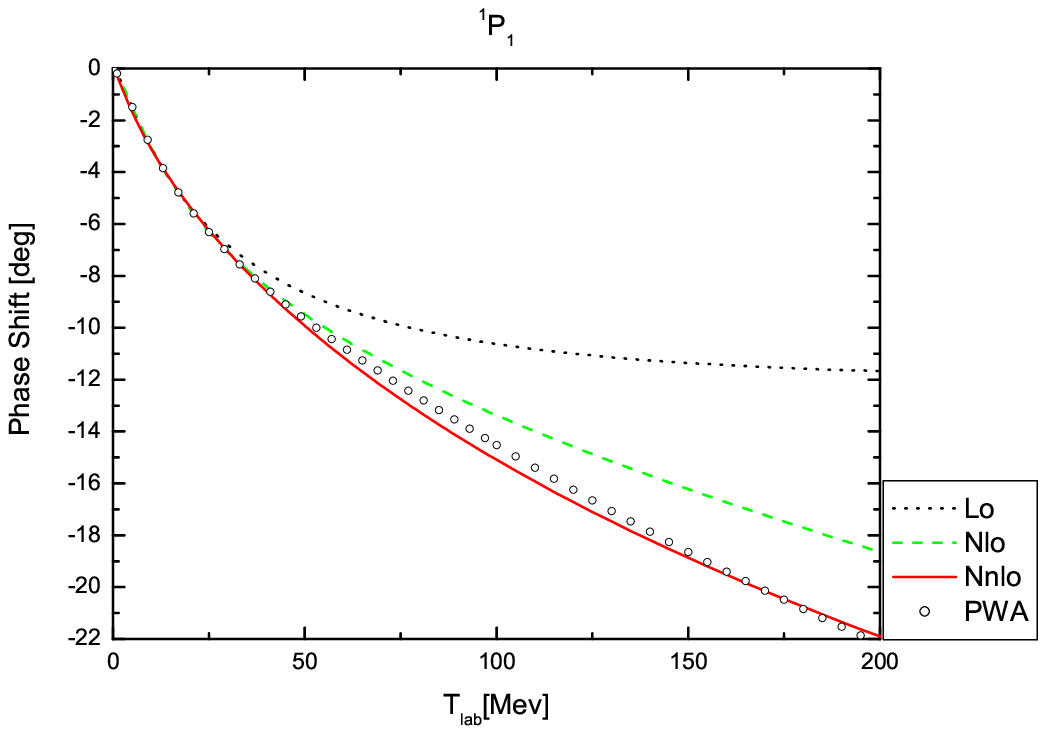}} &
   \resizebox{90mm}{!}{\includegraphics{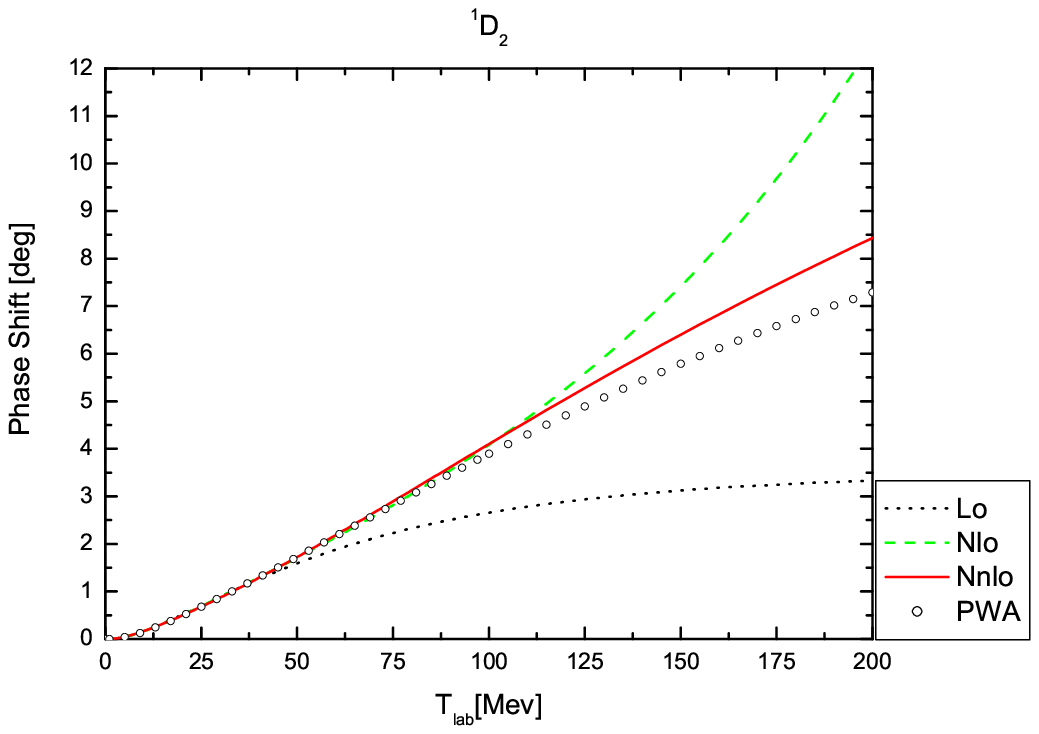}}
\end{tabular} \caption{\footnotesize The predictions
of the phase shifts (versus energy $\text{T}_{\text{lab}}$) for
$^1P_1$ channel (left) and $^1D_2$ channel (right) at LO, Nlo and
Nnlo, against the PWA data. $\text{Re}({\mathcal{G}}_{l})=g_{l;0}$
is fitted at lower energy ends } \label{1p1}
\end{center}
\end{figure}

Following the standard practice\cite{Epel,Rho,KSW,BBSvK,Entem,VA},
we use the PWA data\cite{nij} as our targets. Surprisingly, for
some diagonal channels, e.g., $^1P_1$ and $^1D_2$, it suffices to
use the simplest expansion: $\text{Re}({\mathcal{G}}_{l})\approx
g_{l;0}$. The results for the potential truncated at the first
three chiral orders (leading order(LO), next-to-leading order
(NLO) and next-to-next-to-leading order (NNLO), respectively) are
presented in Fig.\ref{1p1}. The potential at different orders
could be obtained from EGM\cite{Epel}.

It is clear that the predictions of the phase shifts, after
fitting out $\text{Re}({\mathcal{G}}_{l=1,2})=g_{l=1,2;0}$,
improve significantly as the chiral order for the constructed
potential increases (LO, NLO and NNLO). Compared to the results
obtained by the conventional approaches\cite{vK,Epel}, our results
using the simple treatment (coarse as it seems) are fairly
satisfactory, implying that this seemingly coarse approach
contains substantial physical contents of the renormalized
nucleon-nucleon behavior in the low energy regions. In other
words, the efficiency of Pad\'e parametrization of the
$\text{Re}({\mathcal{G}})$ factor to study the nonperturbative
renormalization of the nucleon-nucleon scattering within the EFT
framework is essentially justified, Pad\'e expansion of
$\text{Re}({\mathcal{G}})$ does capture the essences of the
nonperturbative feature of $T$-matrix. The next steps are to apply
the strategy based on the Pad\'e expansion of the
$\text{Re}({\mathcal{G}})$ factor in other channels (including the
more interesting coupled channels) and other issues. These works
are in progress.

In summary, we have discussed the nonperturbative renormalization
of the NN scattering in a compact parametrization of the
$T$-matrix. The main points are explicated in a simple treatment
of the ${\mathcal{G}}$ factor of this compact parametrization
based on the Pad\'e expansion. The efficiency of this seemingly
coarse but simple approach was demonstrated  in two diagonal
channels, $^1P_1$ and $^1D_2$.

\section*{Acknowledgement}
JFY wishes to thank Dr. E. Ruiz Arriola and Dr. J. Gegelia for
helpful communications on the topic. This project is supported in
part by the National Natural Science Foundation under Grant No.
10205004.

\end{document}